\documentclass[conference]{IEEEtran}
\IEEEoverridecommandlockouts

\usepackage{algorithm}
\usepackage{algorithmic}
\def\BibTeX{{\rm B\kern-.05em{\sc i\kern-.025em b}\kern-.08em
    T\kern-.1667em\lower.7ex\hbox{E}\kern-.125emX}}

\usepackage{url}
\usepackage[utf8]{inputenc}
\usepackage[T1]{fontenc}
\usepackage{fancyvrb}
\usepackage{mathtools}
\usepackage{listings}
\usepackage{mathrsfs}
\usepackage{amsbsy}
\usepackage[pdftex]{graphicx}
\usepackage{color}
\usepackage{graphicx}
\usepackage{caption}
\usepackage{subcaption}
\usepackage{picinpar}
\usepackage{lipsum}
\usepackage{wrapfig}
\usepackage{mathtools}
\usepackage{listings}
\usepackage{multirow}% http://ctan.org/pkg/multirow
\usepackage[draft,nosilent,margin]{fixme}
\usepackage{breqn}
\usepackage[figuresright]{rotating}
\usepackage{booktabs}
\usepackage{pgfplots}
\usepackage{tikz}
\usepgfplotslibrary{external,fillbetween,statistics}
\usepackage{tabu}
\usetikzlibrary{matrix,positioning,fit,fadings}
\pgfplotsset{compat=newest}
\usetikzlibrary{patterns}
\usetikzlibrary{pgfplots.groupplots}
\usetikzlibrary{mindmap}
\usepackage{microtype}
\usepackage{times}
\usepackage{fullpage}
\usepackage{color, colortbl}
\usepackage{verbatimbox}
\usepackage{kantlipsum}
\usepackage[breakable, theorems, skins]{tcolorbox}
\makeatletter
\newcommand\resetstackedplots{
\makeatletter
\pgfplots@stacked@isfirstplottrue
\makeatother
\addplot [forget plot,draw=none] coordinates{(1,0) (2,0) (3,0)};
}

\newboolean{showcomments}
 \setboolean{showcomments}{true}
 \ifthenelse{\boolean{showcomments}}
 { \newcommand{\mynote}[3]{
    \fbox{\bfseries\sffamily\scriptsize#1}
    {\small\textsf{\emph{\color[HTML]{#3}{#2}}}}\\}}
 { \newcommand{\mynote}[3]{}}

\definecolor{listinggray}{gray}{0.9}
\definecolor{lbcolor}{rgb}{0.9,0.9,0.9}
\newcommand*\circled[1]{\tikz[baseline=(char.base)]{%
            \node[shape=circle,fill=gray!20,draw,inner sep=.7pt] (char) {#1};}}

\begin{document}

\title{Network in Disaggregated Datacenters}

\author{%
    \IEEEauthorblockN{%
     \parbox{\linewidth}{\centering
        Brice Ekane$^1$, Yohan Pipereau$^2$, Boris Teabe$^3$, Alain Tchana$^{4,5}$,
    Gael Thomas$^2$, Noel de palma$^1$,\\ Daniel Hagimont$^3$
     }%
    }%

    \IEEEauthorblockA{\textit{%
    \parbox{\linewidth}{\centering
        $^1$University Grenoble Alpes\\
            $^2$Télécom SudParis\\
            $^3$University of Toulouse\\
            $^4$ENS Lyon\\
            $^5$Inria
            }%
            }%
        }
}

\maketitle

\begin{abstract}

Nowadays, datacenters lean on a computer-centric approach based on monolithic servers which include all necessary hardware resources (mainly CPU, RAM, network and disks) to run applications. Such an architecture comes with two main limitations: (1) difficulty to achieve full resource utilization and (2) coarse granularity for hardware maintenance.

Recently, many works investigated a resource-centric approach called disaggregated architecture where the datacenter is composed of self-content resource boards interconnected using fast interconnection technologies, each resource board including instances of one resource type. The resource-centric architecture allows each resource to be managed (maintenance, allocation) independently.

LegoOS is the first work which studied the implications of disaggregation on the operating system, proposing to disaggregate the operating system itself. They demonstrated the suitability of this approach, considering mainly CPU and RAM resources. However, they didn't study the implication of disaggregation on network resources.

We reproduced a LegoOS infrastructure and extended it to support disaggregated networking. We show that networking can be disaggregated following the same principles, and that classical networking optimizations such as DMA, DDIO or loopback can be reproduced in such an environment. Our evaluations show the viability of the approach and the potential of future disaggregated infrastructures. 
\end{abstract}

\maketitle

\section{Introduction}
\label{introduction}
\begin{figure}
    \centering
\includegraphics[width=.7\columnwidth]{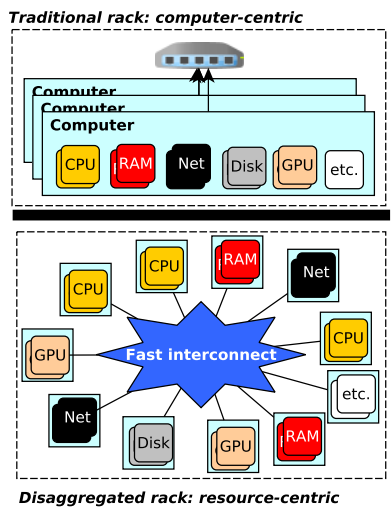}
    \caption{(top) Computer- and (bottom) resource-centric architectures.}
    \label{fig:traditionalVSdisaggregation}
\end{figure}

Nowadays, data centers (DC) lean on a \emph{computer-centric} architecture (see Fig.~\ref{fig:traditionalVSdisaggregation} top), that is, the deployment unit is a \emph{monolithic server} which includes all necessary hardware resources (mainly CPU, RAM, network and disks) to run applications.
Several research works~\cite{Barroso:2007:CEC:1339817.1339894,Delimitrou:2014:QRQ:2654822.2541941, disagreggated2,disagreggated1} have underlined the two main limitations of this architecture.
The first limitation is the difficulty to achieve full resource utilization due to the fact that a process or virtual machine is constrained to a single server boundaries.
The second limitation is the fact that with monolithic servers, adding, removing or changing a single hardware component often involve changing the whole server, which is costly and delays the time to adopt new hardware technologies.
Moreover, the failure of a single component may cause the downfall of the entire server.
These limitations led academia and the industry to imagine a new versatile disaggregated architecture which is \emph{resource-centric} (see Fig.~\ref{fig:traditionalVSdisaggregation} bottom) in which the DC is composed of self-content resource boards interconnected using fast interconnection technologies (e.g., silicon photonics~\cite{mellanox}, infiniband~\cite{intelTCO}).
Each resource board includes instances of one resource type and technology.
The resource-centric architecture allows each resource to evolve independently.
For practicability purposes, DC disaggregation is always studied at the rack scale.

Shan et al.~\cite{Shan:2018:LDD:3291168.3291175} (best paper awards OSDI 2018) studied for the first time the implications of disaggregation on the operating system (OS).
The authors argued that the OS should be disaggregated in the same way as the hardware, resulting to what they called the \emph{split-kernel} model: kernel services are decoupled into loosely-coupled components (called monitors) following the strict separation of concerns~\cite{Shan:2018:LDD:3291168.3291175}.
Each component is responsible for the management of a single resource board type (e.g., CPU) and its code base does not include stuff for managing another component type.
For instance, the monitor of the CPU should not include memory virtualization (e.g., paging) code.
The authors released \emph{LegoOS}, the first split-kernel-based OS.
In its current state, LegoOS takes into account three resource types: CPU, memory, and storage.
Network resources for intra-rack (a kind of loopback, since the rack is seen as a giant computer), inter-rack (within the same DC) and communication with the rest of the world were not considered by the authors.

In this paper, we fill this gap as network is of crucial importance in todays DC, most applications (e.g., machine learning, web servers, databases, etc.) being distributed and accessed through the Internet.
Applying the split-kernel model to CPU, memory and storage components allowed demonstrating the relevance of the disaggregation and split-kernel approaches, but applying these approaches to networking is a much tricky issue. In LegoOS, CPU capacities are managed by monitors called \emph{pComponents} which embed a set of processors and a memory cache (an extra level cache). Memory capacities are managed by monitors called \emph{mComponents} which embed memory banks. Logically, applying dissagregation and split-kernel leads to embedding a set of network devices in a network component (\emph{nComponent}).

However, networking is particular as it heavily involves the CPU and memory components, thus leading to difficult issues and important design choices:

\begin{itemize}
\item First, it seems natural to embed device drivers in nComponents. But what about the protocol stack which can be viewed as a program running in a pComponent (using memory from a mComponent and drivers from nComponents), or embedded in nComponents ?
\item Second, regarding memory management, a sent or received message is read/written from/to memory (in the mComponent) by the device driver (in the nComponent). In a monolithic server, hardware features are integrated for IO memory optimization, such as DMA (Direct Memory Access)~\cite{dma} or DDIO (Data Direct IO)~\cite{ddio}. How do such optimizations translate in a disaggregated architecture ?
\item In a monolithic server, local (to the server) communications are optimized and don't involve a full protocol stack traversal. Is it possible to optimize local communications between two processes within the same pComponent or between two pComponents within the same LegoOS machine ?
\end{itemize}

In this paper, we describe an extension to the LegoOS split-kernel-based system in a disaggregated architecture, aiming at integrating network support. We analyse and motivate the design and integration of networking nComponents in the LegoOS architecture and show that :
\begin{itemize}
\item networking should be disaggregated as nComponents embedding the NIC device drivers as well as the protocol stacks
\item DMA like optimization can be implemented for in memory data transfers in this disaggregated network architecture
\item DDIO like optimization can be implemented for in cache data transfers in this disaggregated network architecture
\item local (intra pComponent or inter pComponents) communications can be optimized as well
\end{itemize}

We reproduced an experimental LegoOS disaggregated platform as described in~\cite{Shan:2018:LDD:3291168.3291175}, implemented in this platform network support following the previous design choices and evaluated the performance of the prototype, demonstrating the suitability of our design.

The rest of the paper is structured as follows. Section~\ref{background} introduces the background information related to disaggregation and the LegoOS system. Section~\ref{networkingInDisaggregatedRack} presents the integration of networking in this environment. Section~\ref{evaluations} reports on our performance evaluation of this prototype. After a review of related works in Section~\ref{rw}, we conclude in Section~\ref{conclusion}.

\section{Background}
\label{background}
This section first introduces the main concepts and the system that we use.

\subsection{Disaggregation}
\label{disaggregation}
Disaggregation splits the monolithic computer into a number of single-resource boards that communicate using a fast interconnect (e.g., silicon photonics \cite{Glick:18}), see Fig.~\ref{fig:traditionalVSdisaggregation}.
Each resource board includes only one resource type (e.g., CPU, main memory, storage or network).
Therefore, the rack is seen as a ``giant'' computer.
An application running in such a disaggregated rack uses several resource boards at a time (at least one CPU and one memory board).
The main benefits expected from disaggregation are as follows.
\emph{(1) It facilitates the scalability of each resource type.}
For instance, the memory capacity of the rack can be increased without increasing the number of CPUs.
\emph{(2) It facilitates application scalability} since the application can use the entire resources of the rack.
This reduces the development burden (of implementing a distributed application) for the application owner.
\emph{(3) It allows optimal resource utilization} (thus good energy proportionality) by minimizing resource fragmentation (with monolithic servers, resource usage is often unbalanced, e.g. memory is fully used and CPUs are under-used).
\emph{(4) It limits the failure impact of a given resource type.}
The failure of a blade is limited to that blade and it does not impact other blades neither of the same type or of a different type.
\emph{(5) It facilitates the rapid integration of new technologies.}
For instance, the integration in the rack of a new CPU generation does not require to buy a whole computer, and thus other resource types.
All these advantages make us and many others~\cite{LimCMRRW09,Lim:2012:SID:2192603.2192683,Shan:2018:LDD:3291168.3291175} believe that next generation DC will be disaggregated.

Disaggregation is still in its early stages, both on the hardware and software sides.
Several manufacturers have started proposing intermediate approaches for rack disaggregation based on micro computers, for example Intel Rack-Scale, AMD SeaMicro~\cite{amd} and HP Moonshot~\cite{hp}.
However, they cannot solve all the issues of monolithic computers since their hardware model is still a monolithic one.
dRedBox \cite{dredbox} is, to the best of our knowledge, the only open project trying to roughly decouple hardware resources.
At the software level, a recent breakthrough is \emph{LegoOS} \cite{Shan:2018:LDD:3291168.3291175}, which introduces the \emph{split-kernel} model as a way of building OSes for disaggregated racks.

\subsection{The split-kernel model and LegoOS}
\label{splitKernel}
In a monolithic server, the OS runs on a single board and its code base is monolithic in the sense that it includes the code to manage all resource types.
The OS assumes local access to all resources.
Shan et al. \cite{Shan:2018:LDD:3291168.3291175} introduced an OS blueprint for disaggregated racks called \emph{split-kernel}.
The latter consists in designing the OS of a disaggregated rack as a disaggregated OS composed of several specialized and loosely-coupled specialized OSes called \emph{monitors}.
Each monitor runs at and manages a unique board.
It operates for its own functionality and communicates with other monitors when there is a need to access other boards' resources.
Shan et al. \cite{Shan:2018:LDD:3291168.3291175} is the first research group which proposed an OS design for real disaggregated racks.
They prototyped the split-kernel model in \emph{LegoOS}.

\begin{figure}
    \centering
\includegraphics[width=.7\columnwidth]{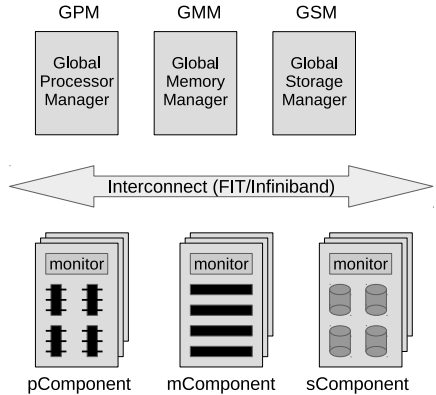}
	\caption{Architecture of LegoOS.}
	\label{fig:archiLegoos}
\end{figure}

Fig.~\ref{fig:archiLegoos} presents an overview of LegoOS' architecture.
Its current design targets three resource types: processor, memory, and storage, embeded in \emph{pComponents}, \emph{mComponents}, and \emph{sComponents} respectively.
LegoOS uses a two-level resource management mechanism.
At the higher level, it uses global resource managers (noted \emph{GPM}, \emph{GMM} and \emph{GSM}) to perform coarse-grained global resource allocation and load balancing for processor, memory and storage respectively. Such managers run on any pComponent of the LegoOS machine.
They only maintain approximate resource usage and load information.
At the lower level, each resource board is locally managed by a monitor.
Each monitor employs its own policies and mechanisms to manage its local resources.
Shan et al. \cite{Shan:2018:LDD:3291168.3291175} mainly described the implementation of pComponent and mComponent.

LegoOS moves all hardware memory functionalities to mComponents (e.g., page tables, TLBs) and leaves only caches at the pComponent side.
Therefore, pComponents only see virtual addresses, which are thus used to access caches.
LegoOS assumes that each pComponent includes a larger (e.g., 1 GB) Last-Level-Cache (called \emph{exCache}) to entirely host both its monitor and a significant portion of applications working set, thus minimizing accesses to mComponents.
Each mComponent can choose its own memory allocation technique and virtual to physical memory address mappings (e.g., segmentation).

Since there is no real disaggregated rack, Shan et al. \cite{Shan:2018:LDD:3291168.3291175} emulated disaggregated boards in LegoOS using commodity servers by limiting their internal hardware usages.
To emulate mComponents, the number of usable cores of the server is limited to two while the emulation of pComponent uses all cores and exploits the main memory of the server as exCache.
The communication between boards is emulated using Mellanox Infiniband adapters, switches and links.
Each monitor embeds a customized RDMA-based RPC framework called \emph{FIT}~\cite{Shan:2018:LDD:3291168.3291175} which eases the utilization of RDMA.

We have faithfully reproduced LegoOS experimental environment in our lab in order to study the integration of network boards, enabling intra- and inter-rack communication, and communication with the rest of the world.

\section{Networking in a disaggregated datacenters}
\label{networkingInDisaggregatedRack}
Our contribution involves adding a network component (hereafter \emph{nComponent}) to the disaggregated rack following the split-kernel model.
nComponent allows processes within the disaggregated rack to communicate both with each other and with the rest of the world using network primitives (e.g., socket BSD API).
To integrate nComponent, we need to answer four main questions:
What is the composition of nComponent from a hardware perspective?
How can essential and novel hardware features such as DMA and DDIO be adapted to the disaggregated paradigm?
What are the services of the nComponent monitor?
How to provide best performance?

\subsection{Hardware design}
\label{hardwareComposition}
\begin{figure}
\centering
\includegraphics[width=.9\columnwidth]{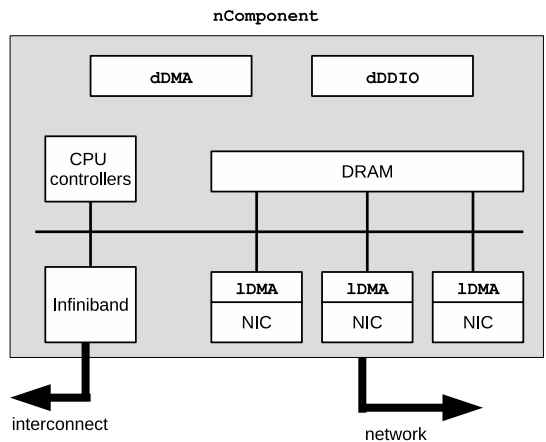}
\caption{Hardware architecture of nComponent}
\label{fig:archiNcomp}
\end{figure}

Fig~\ref{fig:archiNcomp} presents the hardware architecture of a nComponent.
\texttt{NICs} are either classical or sophisticated (smart NICs~\cite{smartNics}) network adapters which include packet reception and transmission circuits, and can be referenced using MAC and IP addresses.
They connect the disaggregated rack with other network devices in the data center (e.g., switches).
According to the split-kernel model, nComponent as well as other disaggregated devices, includes a set of \texttt{controllers} (comparable to CPUs) to run the network monitor.
The binary code of the latter resides in a local memory (\texttt{DRAM} in the figure), instead of a remote mComponent for performance purposes.
The local memory also hosts incoming and outgoing packets to/from NICs.
This is achieved by local DMA (noted \texttt{lDMA} in the figure) engines integrated in NICs.
For all these reasons, the local memory of nComponent must have a substantial size, in the range of giga bytes.

The nComponent also includes a disaggregated DMA (noted \texttt{dDMA}) engine and a disaggregated DDIO (noted \texttt{dDDIO}) engine which implement packet transfer with mComponents and pComponents' extCache respectively. Although DMA and DDIO are implemented in the hardware in monolithic servers, they are implemented by the software monitor in the nComponent.

\subsection{Software design overview}
\label{generalOverview}
\begin{figure}
\centering
\includegraphics[width=1\columnwidth]{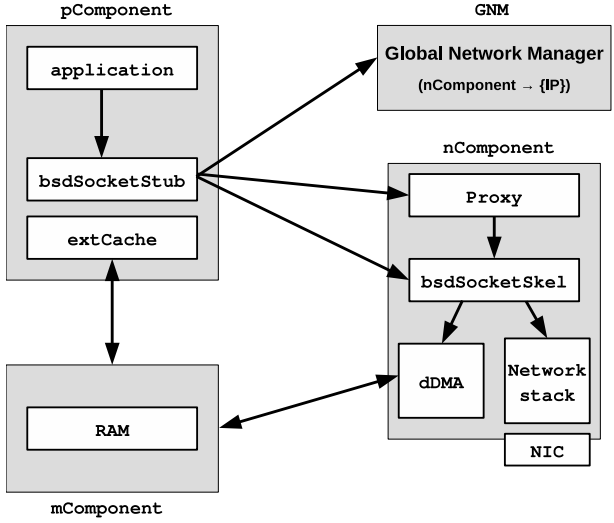}
\caption{General overview of LegoOs extended with nComponent}
\label{fig:generalOverview}
\end{figure}

Fig~\ref{fig:generalOverview} presents the extended architecture of LegoOS which takes into account networking aspects.
Our contribution in this architecture is threefold.
\emph{(1)} The Global Network Manager (\texttt{GNM}) runs on the same machine as the other LegoOS managers (see Section~\ref{background}).
Once started, GNM initializes a list of available nComponents within the rack, including the IP address associated with each NIC, as it implements NIC allocation policy to processes.
\emph{(2)} The pComponent monitor is extended with a \texttt{bsdSocketStub} object which provides a complete BSD socket API interface to processes.
Following the separation of concerns behind the split-kernel model, the implementation core of the socket BSD API is made inside nComponent.
It is provided by a \texttt{bsdSocketSkeleton} object.
Each process in a pComponent performing a network operation is allocated an instance of \texttt{bsdSocketStub}, which manages the socket state and forwards all calls to its counterpart \texttt{bsdSocketSkeleton} instance in the nComponent, implementing a RPC like mechanism.
\emph{(3)} nComponent runs the network monitor, an operating system dedicated to NIC management.
It includes \texttt{bsdSocketSkeleton} instances, a \texttt{Proxy} (it is the entry point for the nComponent allowing to create \texttt{bsdSocketSkeleton} instances for communications), network routing rules, NIC drivers and a setup.
The latter initializes the nComponent internal state (e.g., IP addresses associated with NICs) and registers the nComponent and its IPs in the GNM.

The next sections detail the implementation of the main steps of connection establishment and message exchange.

\subsection{Connection establishment and termination (\texttt{socket, bind, connect})}
\label{connectionEsatablishment}
The invocation of \emph{socket()}, which is the first call performed by a network process in a pComponent, instantiates a \texttt{bsdSocketStub}  instance whose role is to handle all socket BSD calls.
When \emph{bind()} is invoked, \texttt{bsdSocketStub} sends to GNM the (local) IP address parameter of the \emph{bind()} operation, asking the GNM the nComponent associated with this IP address (GNM knows the IP addresses associated with each nComponents). The nComponent identifier is returned to \texttt{bsdSocketStub} which registers it.
 \texttt{bsdSocketStub} then connects with the Proxy on the target nComponent, and requests the creation of a \texttt{bsdSocketSkeleton} instance, and sends to it all information related to the created socket.
These informations are: basic socket information (socket type, protocol, IP address, port), pComponent and \texttt{bsdSocketStub} instance identifiers, allowing the \texttt{bsdSocketSkeleton} instance to send messages back to the \texttt{bsdSocketStub} instance. The \texttt{bsdSocketSkeleton} instance registers these informations and creates in the nComponent a local socket on the protocol stack.
From then on, the \texttt{bsdSocketStub} communicates directly with this nComponent/\texttt{bsdSocketSkeleton} using a message-based protocol (similar to a RPC, but invocations can be performed in both directions).
A call to \emph{accept()} is forwarded by \texttt{bsdSocketStub} to \texttt{bsdSocketSkeleton} which reproduces it on the local socket and blocks waiting for a connection request, which (when received) is forwarded by \texttt{bsdSocketSkeleton} back to \texttt{bsdSocketStub}.
Similarly, a call to \emph{connect()} is forwarded by \texttt{bsdSocketStub} to \texttt{bsdSocketSkeleton} and reproduced on the local socket.

\subsection{Packet transmission (\texttt{send})}
\label{packetTransmission}
\begin{figure}
\centering
\includegraphics[width=1\columnwidth]{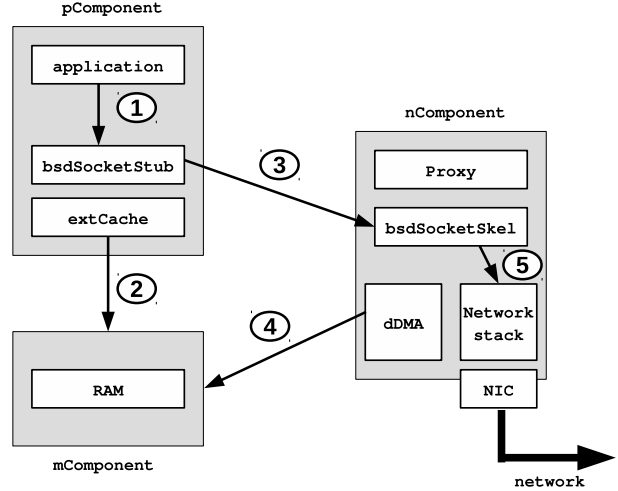}
\caption{Packet transmission step}
\label{fig:networkTransmission}
\end{figure}

We consider hereafter that the default version relies on DMA access to memory. We describe in Section~\ref{ddio} the DDIO based communication scheme.
Fig~\ref{fig:networkTransmission} summarizes the implementation steps of packet sending, issued by a process on the pComponent.
On \texttt{send()} invocation (step \circled{1}), \texttt{bsdSocketStub} transfers the data to be sent from \textit{extCache} (on pComponent) to mComponent (step \circled{2}). This is a kind of data flush from extCache to memory.
It then informs (step \circled{3}) the \texttt{bsdSocketSkeleton} of the presence of data to be sent over the network.
To this end, \texttt{bsdSocketStub} provides to \texttt{bsdSocketSkeleton}
the size of the data to be sent and its memory address in mComponent.
Upon receiving this order, \texttt{bsdSocketSkeleton} configures dDMA in such a way that data to be sent can be loaded from mComponent and stored in DRAM (\circled{4}) when accessed by the network stack.
Subsequently, \texttt{bsdSocketSkeleton} invokes the \texttt{send()} function on the local socket (step \circled{5}), walking through the kernel network stack and transfering data to the NIC driver. Notice that both dDMA and lDMA are involved when the network stack effectively handles the data.

\subsection{Packet reception (\texttt{recv})}
\label{packetReception}

When a packet arrives at the NIC, the lDMA engine copies the packet to the nComponent's DRAM and the data is registered in the \texttt{bsdSocketSkeleton}.
When \texttt{recv()} is invoked in the pComponent, the call is forwarded from \texttt{bsdSocketStub} to \texttt{bsdSocketSkeleton} (with the address of the buffer in mComponent), which asks dDMA to copy the read data from the nComponent's DRAM to the buffer in mComponent. Then, \texttt{bsdSocketSkeleton} responds to \texttt{bsdSocketStub} with the size of the read data. The reading process in the pComponent will then access the data, loading it from mComponent to extCache.

\subsection{Disaggregated DDIO}
\label{ddio}
Data Direct I/O Technology (Intel DDIO) is a feature introduced in Intel processes that allows a NIC to directly exchange data with the CPU cache without going through the RAM.
This considerably increases the network speed, reduces latency and finally reduces energy consumption.
DDIO in a disaggregated environment consists of allowing nComponent to send/receive packets directly to/from pComponent (its extCache) without going through mComponent.
To make this possible in our design, the workflow of packet transmission and reception presented above slightly changes as follows.
On transmission (\texttt{send()}), \texttt{bsdSocketStub} is able to transmit the data to be sent directly from pComponent's extCache to nComponent's DRAM (assuming the data resides in the cache).
On reception \texttt{recv()}, \texttt{bsdSocketSkeleton} is able to transmit the received data directly to the extCache (in an extry associated with the reception buffer in memory).

\subsection{Optimizations}
\label{optimizations}
The implementation presented above strictly follows the split-kernel model, that is said separation of concerns:
nComponent and GNM are the only components containing network services.
This has a severe consequence for intra-rack communications (e.g., which use the \textit{loopback} interface).
In fact, all network communications, including intra-rack communications, systematically involve nComponent.
This is not necessary when the two processes are within the same disaggregated rack.
This limitation increases intra-rack communication latency in several use cases.
It is important to note that in a disaggregated rack, the probability to see two pieces of the same application colocated atop the same rack (thus communicating) is very high compared to traditional monolithic servers.
This is because the rack has a bigger size than a server.
We propose an optimization which requires the distortion of the split-kernel model for tackling this limitation.

Our optimization reduces intra- and inter-pComponent communication inside the same disaggregated rack.
When \emph{connect()} is invoked on \texttt{bsdSocketStub}, the stub asks the GNM whether the target IP address is associated with a nComponent in the rack. If so, the connection is local to the rack and optimization can take place. Then, \texttt{bsdSocketStub} invokes the Proxy on that nComponent to ask the location of the process which made a bind on the target IP/port (the target pComponent and \texttt{bsdSocketStub} instance identifiers). Thus, a direct connection between the origin and target \texttt{bsdSocketStubs} can be established.
If the communication is intra-pComponent, the two \texttt{bsdSocketStubs} will communicate using Unix Pipe.
If the communication is inter-pComponent, the two \texttt{bsdSocketStub} will communicate using FIT (LegoOS interconnect communication protocol).
This optimization requires that \texttt{bsdSocketStub}, which is part of pComponent monitor, includes core networking code.
The authors of LegoOS acknowledged a similar distortion to the application of the split-kernel model for achieving optimal performance in the relationship between the CPU and memory (leading to extCache extra cache level).

\section{Evaluations}
\label{evaluations}

\begin{footnotesize}

 \begin{table*}[ht]
 \begin{center}

 \begin{tabular}{|c|c|c|c|}
 \hline
 CPU & Memory size & Ethernet card  & Infiniband card \\
 \hline
 Intel(R) Core(TM) i7-3770  &  4 GB & Intel Corporation 82574L Gigabit  &  Mellanox MT27500\\
 \hline
 Intel(R) Xeon(R) CPU E5-1603 v3  & 8 GB &  Intel Corporation 82574L Gigabit &  Mellanox MT27500 \\
 \hline
 Intel Core Processor (Haswell, no TSX) & 3 GB &  Intel Corporation 82574L Gigabit &  Mellanox MT27500\\
 \hline

 \end{tabular}
  \end{center}

    \caption{Servers characteristics}
 \label{servers}
 \end{table*}
 \end{footnotesize}
This section presents the performance evaluations. We aim at answering the following questions:
\begin{itemize}
\item What is the network latency and throughput with our network stack implementation in a disaggregated rack?
\item What is the overhead with a disaggregated rack compared to a monolithic Linux system?
\item What performance can we expect with a real world benchmark?
\end{itemize}
The goal of the comparison with a monolithic Linux system is to show that a disaggregated server can perform in the same order of magnitude as a monolithic server, and that a real disaggregated hardware (especially a dedicated interconnect instead of Inifiniband) would lead to a reasonable overhead while benefiting from the flexibility of disaggregation.

To this end, we rely on two benchmarks for evaluation: a micro-benchmark which is a TCP client/server application and a macro-benchmark which is a spark streaming like benchmark (representing a real world benchmark).  The micro-benchmark is used to evaluate the internal mechanisms of our implementation while the macro-benchmark provides a global performance view.

We carried out the evaluations with 3 configurations:
\begin{itemize}
 \item \textbf{DMA (default configuration)}.  This configuration corresponds to the basic implementation (with dDMA) of our network stack on LegoOS with no optimization. Like the traditional DMA feature, it limits the intervention of CPU (here pComponent) in data transfers, the dDMA (in the nComponent) being responsible for fetching data from the mComponent. This is better suited for large data transfers (where throughput is important) and it offloads processing from the pComponent to the nComponent.
  \item \textbf{DDIO}. This corresponds to the optimized implementation similar to the intel CPU DDIO feature. This mechanism involves CPU (pComponent) for each data transfer and is better suited for small data transfers (where latency is important), in the extCache and involving limited CPU.
  \item \textbf{Linux}. This corresponds to evaluation with a standard Linux system (without LegoOS).
\end{itemize}

\subsection{Hardware setup and LegoOS deployment}
Since there is no real resource disaggregation hardware, we emulated our disaggregated environment using 5 standard servers, 5 Infiniband network adapters and an Infiniband switch. Table~\ref{servers} presents our server characteristics. Each server has an InfiniBand network adapter (\emph{Mellanox Technologies MT27500 Family [ConnectX-3] }) and there are all connected through a Mellanox  Infiniband switch (\emph{Mellanox IS5022 Infiniscale-IV 8-ports}). Notice that the heterogeneity of the environment is typical to a disaggregated rack which will have pComponent, mComponent and nComponent from different vendors. In our emulated environment, servers are used as components by  limiting their internal hardware usages, i.e. a server considered as a pComponent will see its CPU used but not its network adapter, while a server considered as a nComponent will have its network adapter used but only few CPU for its internal monitor. We used 2 servers as pComponent with a exCache of size 1 GB each, 1 servers as nComponents with 1 GB of DRAM, 1 servers as mComponent with 8 GB of RAM and finally 1 server as sComponent.

The Mellanox cards used in our infrastructure are identical to those used by the designers of LegoOs~\cite{Shan:2018:LDD:3291168.3291175}. Thus, the latency is similar to that contained in the article which is of the order of 8 $usec$ for message round-trip.

We downloaded the current available version of LegoOS from git~\cite{git} and deployed it in our infrastructure.

\begin{figure*}[h!]
\centering
\includegraphics[width=1.9\columnwidth]{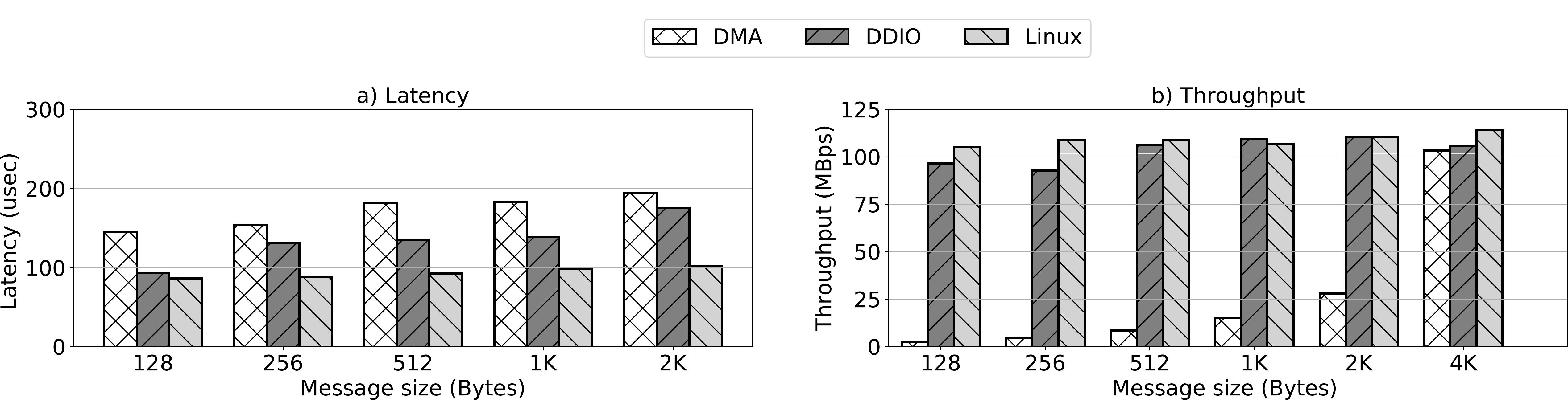}
\caption{High level metrics for micro-benchmark evaluation (latency and throughput) on LegoOS.}
\label{fig:latency}
\end{figure*}

\subsection{Micro-benchmark evaluation}
\label{micro}

\textbf{Basic evaluation.} We use a simple TCP client/server application as micro-benchmark. The experimental procedure is as follow: the server is deployed on the system to be evaluated while the client is on a remote machine running a standard Linux. The client sends and receives a flow of fixed size messages to/from the server. This procedure is repeated with various message sizes from 128 to 2048 Bytes. Fig~\ref{fig:latency} presents the results with the latency and the throughput metrics. The first figure (a) presents latency results while the last figure plots the throughput (b).

What immediately gain our attention is the high latency of DMA compared to that of DDIO and Linux (on Fig~\ref{fig:traditionalVSdisaggregation} (a)). This is simply explained by the additional communications between components with DMA (the packet must transit through the mComponent before moving to nComponent), while with DDIO there is a direct communication between the pComponent and nComponent. As previously explained, DDIO is better suited for small data transfers where latency is more important and DDIO's performance are in the same order of magnitude as Linux for messages smaller than 1K.

Regarding the throughput, the DDIO and Linux performances are quite close. However, DDIO is not best suited for large data transfer as it heavily involves the pComponent. We can observe that the throughput for DMA is much lower than that of Linux and DDIO. This is especially true for small message sizes as it involves many interactions with the mComponent. However, for larger message sizes (we extended the experiment with 4K messages), DMA throughput comes close to Linux (as less interactions with mComponent are needed).
Notice that messages are fetched on demand from the mComponent and we did not implement message packing which would limit interactions with the mComponent (this is a track of improvement).

\textbf{Connection establishment.} The results presented above do not include connection establishment times which is the time spend when the client invoke the \texttt{connect} syscall. The connection times are very close (255 us for LegoOS and 252 us for Linux) as LegoOS only adds a RPC interaction between the \texttt{bsdSocketStub} in the pComponent and the \texttt{bsdSocketSkeleton} in the nComponent, which in turn runs the same connection opening as Linux.

\textbf{Scalability evaluation.} Regarding scalability, we evaluated the ability of the nComponent to handle an important communication load. We measured the amount of CPU (cores) needed to saturate our 1~Gb Ethernet card, on emission and on reception (the other side being provided enough CPU). This is a means to dimension the maximum number of cores in the nComponent. We observed that 15\% and 25\% of a core were required respectively for emission and reception, reaching an effective bandwidth of 880 Mb/s. We also observed that additional NICs (and cores accordingly) in the nComponent allowed to scale up the nComponent and absorbe a more important load from pComponents.

\begin{figure}[h!]
\centering
\includegraphics[width=.9\columnwidth]{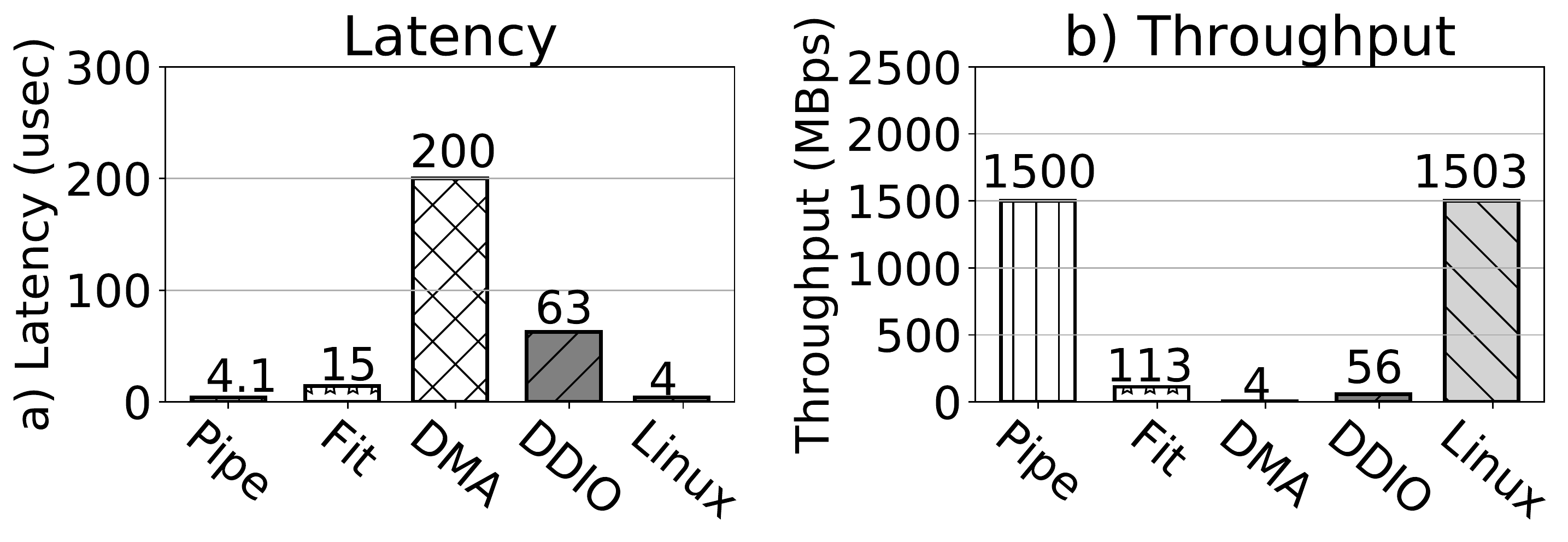}
\caption{Intra and inter-pComponent evaluation. The figures present the latency (lower is better) and the throughput (higher is better).}
\label{fig:localhost}
\end{figure}

\textbf{Inter and intra pComponent communication.} Inter and intra pComponent communication has been optimized as described in Section~\ref{networkingInDisaggregatedRack}. We evaluated the benefits of these optimizations and estimated the performance gain. Fig~\ref{fig:localhost} plots the results obtained for the latency and the throughput. We can observe a huge difference in performance (latency and throughput) between DMA and DDIO (which involve communication with the nComponent), and optimized implementations for both intra (Pipe) and inter-pComponent (Fit) communications. The good performance with the optimization comes from the non-participation of nComponent and mComponent  in the communication process, and the use of respectively pipe and FIT for intra and inter pComponent communications. Therefore, with the optimizations we proposed, performance is close to that obtained with a monolithic Linux system.

\subsection{Macro-benchmark}
\label{redisEval}
\begin{figure}
\centering
\includegraphics[width=.9\columnwidth]{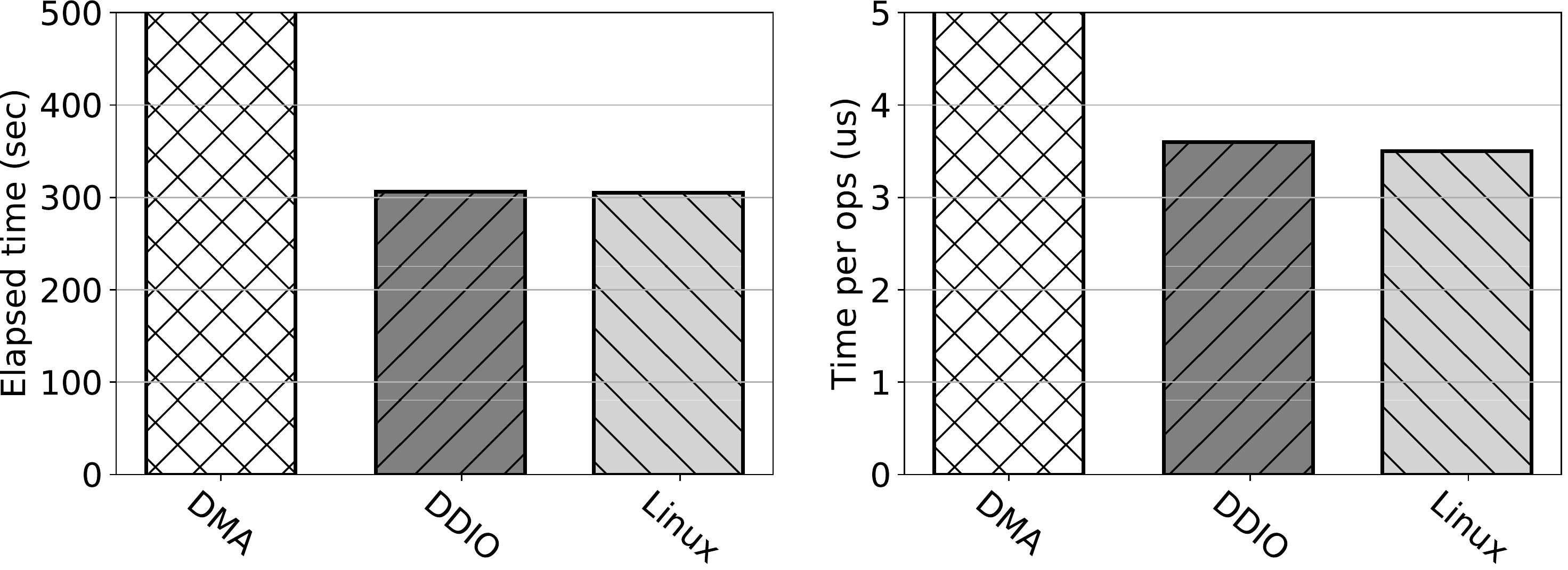}
\caption{Streaming benchmark evaluation}
\label{fig:spark}
\end{figure}

For the macro benchmark evaluation, we used the word count application on a spark-streaming like platform following the map-reduce paradigm. We recorded two metrics: the execution time and the number of ops per seconds (ops/sec). The benchmark consists of a single  mapper and a single reducer: the map runs on the evaluated system (Legos or standard Linux) while the reduce runs on a remote machine running a standard Linux. Therefore, there is intense communication between the evaluated system and the outside  because each pair <word, count>  must be transmitted between the map and the reduce. The processing was done on 1.5 GBytes of data. The results obtained are shown in Figure~\ref{fig:spark}. We can observe that the DDIO version performs in the same time as Linux, since each pair (which is very small) is sent over the network as soon as it is produced. The DMA version cannot reach the same performance level, because its interactions with the mComponent are too penalizing for small data exchanges. Sending larger blocks which don't fit in the extCache would be more favorable for DMA.

\section{Related work}
\label{rw}
In this section, we first review previous works on disaggregation in general, before focussing on OS level works in such environments.
\subsection{Disaggregation}
Disaggregation splits the monolithic computer into a number of resource boards which communicate using a fast interconnect.
This is inspired by various trends from the past such as diskless machines using network storage such as network block devices.
\cite{Lim:2009:DME:1555815.1555789} is the first work on memory disaggregation in datacenters, extending a host memory with remote servers memory, thus decoupling CPU from memory allocations.
In the same vein, several other works have studied memory disaggregation, such as~\cite{Nitu:2018:WZP:3190508.3190537} which proposed enabling remote access to the memory of a suspended server.
Not so long ago were studied fully disaggregated infrastructures, notably with storage rack disaggregation~\cite{Legtchenko:2017:URD:3154601.3154603} and  network requirements for
resource disaggregation~\cite{199305,246900,walraed-sullivan2014theia}. More precisely, \cite{199305} carried out a study to estimate the latency and bandwidth requirements that the interconnect in
disaggregated datacenters must meet to avoid degrading application-level performance with existing network designs. \cite{246900} and \cite{walraed-sullivan2014theia} explored new architecture designs to integrate 1 Tbps silicon photonic, remote nonvolatile memory, and System-on-Chips (SoCs) to datacenter for disaggregation. Although these works are related to networking, they only deal with the hardware level unlike our work which includes a network stack implementation for a disaggregated OS.

Several manufacturers have proposed intermediate approaches for rack disaggregation based on microcomputers. As example we can cite Intel Rack-Scale~\cite{intelrack}, AMD SeaMicro~\cite{amd} and HP Moonshot~\cite{disagreggated1}. All these works correspond to an intermediate step towards full disaggregation.
With the dRedBox European project~\cite{dredbox}, IBM and several academic institutions are currently trying to build a real disaggregated rack prototype, but only focus on hardware aspects.

\subsection{OSes in disaggregation and the splitkernel model}
Several OS architectures try to respond to huge computing capacity environments and scalability issues: Multi-kernelOSes like Barrelfish \cite{10.1145/1629575.1629579} and Helios~\cite{10.1145/1629575.1629597}, multi-instance OSes like Popcorn~\cite{10.1145/2741948.2741962} and Pisces~\cite{10.1145/2749246.2749273}, and finally distributed OSes~\cite{10.1145/354871.354873,10.1145/1067625.806544}.
These OS architectures, even though they look interesting and are designed to handle large infrastructures, are not suitable for disaggregated environments as they don't assume a resource centric infrastructure composed of single resource type boards.
To the best of our knowledge, the only attempt to design an OS for a fully disaggregated infrastructure is LegoOS~\cite{Shan:2018:LDD:3291168.3291175} which introduces the split-kernel model as a
way of building OSes for disaggregated racks. The splitkernel model consists in designing the OS of a disaggregated rack as a disaggregated OS composed of several specialized and loosely-coupled OSes. The authors prototyped their model with \emph{LegoOS} to demonstrate its feasibility. In its current state, LegoOS targets three types of hardware components: CPU, memory, and storage, but does not address network support. In this work we proposed an approach to handle network components in the splitkernel model and we carry out an implementation in LegoOS.

\section{Conclusion}
\label{conclusion}
Shan et al.~\cite{Shan:2018:LDD:3291168.3291175} proposed, for the first time, the split-kernel model as an appropriate operating system design model for disaggregated datacenters.
The authors validated their claim by prototyping the CPU, memory and storage parts of a split-kernel-based operating system.
In the present paper, we studied the support of networking in a disaggregated rack in respect with the split-kernel model.
We showed that this is not a straightforward task because of the strong relationship between network devices and the CPU-memory couple.
In addition, we studied how classical networking features such as DMA, DDIO or loopback can be efficiently taken into account in such an environment.
We extended the original LegoOS prototype and evaluated our implementation using both micro- and macro-benchmarks.
Our evaluations show that with the emulated disaggregated platform defined by LegoOS (that we reproduced), our network support implementation can perform in the same order of magnitude as a monolithic Linux server for latency and very close for bandwidth, leading to very close results for a big data benchmark. Therefore, a real disaggregated hardware (especially a dedicated interconnect instead of Inifiniband) would lead to a reasonable overhead while benefiting from the flexibility of disaggregation.

\bibliography{./paper}

\begin{thebibliography}{10}

\bibitem{amd}
Eamicro, a. amd seamicro sm15000 fabriccompute systems.
\newblock \url{ http://www.seamicro.com/}.
\newblock Accessed: 2020-08-5.

\bibitem{disagreggated1}
Hp moonshot.
\newblock \url{System.http://goo.gl/fteii.}
\newblock Accessed: 2019-09-5.

\bibitem{hp}
Hp moonshot system.
\newblock \url{https://files.vogel.de/vogelonline/vogelonline/files/6284.pdf}.
\newblock Accessed: 2020-08-5.

\bibitem{ddio}
Intel data direct i/o technology (intel ddio).
\newblock \url{
  https://www.intel.com/content/dam/www/public/us/en/documents/technology-briefs/data-direct-i-o-technology-brief.pdf}.
\newblock Accessed: 2020-08-5.

\bibitem{intelrack}
Intel rack scale.
\newblock
  \url{https://www.intel.com/content/dam/www/public/us/en/documents/guides/architecture-spec-v2-4-guide.pdf
  }.
\newblock Accessed: 2020-09-5.

\bibitem{intelTCO}
Intel.intel rack scale architecture:faster service delivery and lower tco..
\newblock
  \url{http://www.intel.com/content/www/us/en/architecture-and-technology/intel-rack-scale-architecture.html.
  }.
\newblock Accessed: 2019-09-5.

\bibitem{git}
Legoos git.
\newblock \url{https://github.com/WukLab/LegoOS}.
\newblock Accessed: 2020-09-5.

\bibitem{mellanox}
Mellanox.rdma aware networks programming usermanual.
\newblock
  \url{.http://www.mellanox.com/related-docs/prodsoftware/RDMAAwareProgrammingusermanual.pdf.
  }.
\newblock Accessed: 2019-09-5.

\bibitem{dma}
Network direct memory access.
\newblock \url{https://patents.google.com/patent/US7836220B2/en }.
\newblock Accessed: 2020-08-5.

\bibitem{disagreggated2}
Seamicro technology overview.
\newblock \url{http://seamicro.com/sites/default/files/SM_TO01_64_v2.5.pdf.}
\newblock Accessed: 2019-09-5.

\bibitem{smartNics}
What is a smartnic.
\newblock \url{https://blog.mellanox.com/2018/08/defining-smartnic/ }.
\newblock Accessed: 2020-08-5.

\bibitem{246900}
K.~Asanovi{\'c}.
\newblock Firebox: A hardware building block for 2020 warehouse-scale
  computers.
\newblock Santa Clara, CA, Feb. 2014. {USENIX} Association.

\bibitem{10.1145/2741948.2741962}
A.~Barbalace, M.~Sadini, S.~Ansary, C.~Jelesnianski, A.~Ravichandran,
  C.~Kendir, A.~Murray, and B.~Ravindran.
\newblock Popcorn: Bridging the programmability gap in heterogeneous-isa
  platforms.
\newblock In {\em Proceedings of the Tenth European Conference on Computer
  Systems}, EuroSys ’15, New York, NY, USA, 2015. Association for Computing
  Machinery.

\bibitem{Barroso:2007:CEC:1339817.1339894}
L.~A. Barroso and U.~H\"{o}lzle.
\newblock The case for energy-proportional computing.
\newblock {\em Computer}, 40(12):33--37, Dec. 2007.

\bibitem{10.1145/1067625.806544}
F.~Baskett, J.~H. Howard, and J.~T. Montague.
\newblock Task communication in demos.
\newblock {\em SIGOPS Oper. Syst. Rev.}, 11(5):23–31, Nov. 1977.

\bibitem{10.1145/1629575.1629579}
A.~Baumann, P.~Barham, P.-E. Dagand, T.~Harris, R.~Isaacs, S.~Peter, T.~Roscoe,
  A.~Sch\"{u}pbach, and A.~Singhania.
\newblock The multikernel: A new os architecture for scalable multicore
  systems.
\newblock In {\em Proceedings of the ACM SIGOPS 22nd Symposium on Operating
  Systems Principles}, SOSP ’09, page 29–44, New York, NY, USA, 2009.
  Association for Computing Machinery.

\bibitem{Delimitrou:2014:QRQ:2654822.2541941}
C.~Delimitrou and C.~Kozyrakis.
\newblock Quasar: Resource-efficient and qos-aware cluster management.
\newblock {\em SIGARCH Comput. Archit. News}, 42(1):127--144, Feb. 2014.

\bibitem{199305}
P.~X. Gao, A.~Narayan, S.~Karandikar, J.~Carreira, S.~Han, R.~Agarwal,
  S.~Ratnasamy, and S.~Shenker.
\newblock Network requirements for resource disaggregation.
\newblock In {\em 12th {USENIX} Symposium on Operating Systems Design and
  Implementation ({OSDI} 16)}, pages 249--264, Savannah, GA, Nov. 2016.
  {USENIX} Association.

\bibitem{Glick:18}
M.~Glick, S.~Rumley, and K.~Bergman.
\newblock Silicon photonics enabling the disaggregated data center.
\newblock In {\em Advanced Photonics 2018 (BGPP, IPR, NP, NOMA, Sensors,
  Networks, SPPCom, SOF)}, page NeM3F.4. Optical Society of America, 2018.

\bibitem{10.1145/354871.354873}
K.~Govil, D.~Teodosiu, Y.~Huang, and M.~Rosenblum.
\newblock Cellular disco: Resource management using virtual clusters on
  shared-memory multiprocessors.
\newblock {\em ACM Trans. Comput. Syst.}, 18(3):229–262, Aug. 2000.

\bibitem{dredbox}
K.~Katrinis, D.~Syrivelis, D.~Pnevmatikatos, G.~Zervas, D.~Theodoropoulos,
  I.~Koutsopoulos, K.~Hasharoni, D.~Raho, C.~Pinto, F.~Espina, and et~al.
\newblock Rack-scale disaggregated cloud data centers: The dredbox project
  vision.
\newblock {\em Proceedings of the 2016 Design, Automation \& Test in Europe
  Conference \& Exhibition (DATE)}, 2016.

\bibitem{Legtchenko:2017:URD:3154601.3154603}
S.~Legtchenko, H.~Williams, K.~Razavi, A.~Donnelly, R.~Black, A.~Douglas,
  N.~Cheriere, D.~Fryer, K.~Mast, A.~D. Brown, A.~Klimovic, A.~Slowey, and
  A.~Rowstron.
\newblock Understanding rack-scale disaggregated storage.
\newblock In {\em Proceedings of the 9th USENIX Conference on Hot Topics in
  Storage and File Systems}, HotStorage'17, pages 2--2, Berkeley, CA, USA,
  2017. USENIX Association.

\bibitem{Lim:2009:DME:1555815.1555789}
K.~Lim, J.~Chang, T.~Mudge, P.~Ranganathan, S.~K. Reinhardt, and T.~F. Wenisch.
\newblock Disaggregated memory for expansion and sharing in blade servers.
\newblock {\em SIGARCH Comput. Archit. News}, 37(3):267--278, June 2009.

\bibitem{Lim:2012:SID:2192603.2192683}
K.~Lim, Y.~Turner, J.~R. Santos, A.~AuYoung, J.~Chang, P.~Ranganathan, and
  T.~F. Wenisch.
\newblock System-level implications of disaggregated memory.
\newblock In {\em Proceedings of the 2012 IEEE 18th International Symposium on
  High-Performance Computer Architecture}, HPCA '12, pages 1--12, Washington,
  DC, USA, 2012. IEEE Computer Society.

\bibitem{LimCMRRW09}
K.~T. Lim, J.~Chang, T.~N. Mudge, P.~Ranganathan, S.~K. Reinhardt, and T.~F.
  Wenisch.
\newblock Disaggregated memory for expansion and sharing in blade servers.
\newblock In S.~W. Keckler and L.~A. Barroso, editors, {\em ISCA}, pages
  267--278. ACM, 2009.

\bibitem{10.1145/1629575.1629597}
E.~B. Nightingale, O.~Hodson, R.~McIlroy, C.~Hawblitzel, and G.~Hunt.
\newblock Helios: Heterogeneous multiprocessing with satellite kernels.
\newblock In {\em Proceedings of the ACM SIGOPS 22nd Symposium on Operating
  Systems Principles}, SOSP ’09, page 221–234, New York, NY, USA, 2009.
  Association for Computing Machinery.

\bibitem{Nitu:2018:WZP:3190508.3190537}
V.~Nitu, B.~Teabe, A.~Tchana, C.~Isci, and D.~Hagimont.
\newblock Welcome to zombieland: Practical and energy-efficient memory
  disaggregation in a datacenter.
\newblock In {\em Proceedings of the Thirteenth EuroSys Conference}, EuroSys
  '18, pages 16:1--16:12, New York, NY, USA, 2018. ACM.

\bibitem{10.1145/2749246.2749273}
J.~Ouyang, B.~Kocoloski, J.~R. Lange, and K.~Pedretti.
\newblock Achieving performance isolation with lightweight co-kernels.
\newblock In {\em Proceedings of the 24th International Symposium on
  High-Performance Parallel and Distributed Computing}, HPDC ’15, page
  149–160, New York, NY, USA, 2015. Association for Computing Machinery.

\bibitem{Shan:2018:LDD:3291168.3291175}
Y.~Shan, Y.~Huang, Y.~Chen, and Y.~Zhang.
\newblock Legoos: A disseminated, distributed os for hardware resource
  disaggregation.
\newblock In {\em Proceedings of the 12th USENIX Conference on Operating
  Systems Design and Implementation}, OSDI'18, pages 69--87, Berkeley, CA, USA,
  2018. USENIX Association.

\bibitem{walraed-sullivan2014theia}
m.~walraed sullivan, J.~Padhye, and D.~Maltz.
\newblock Theia: Simple and cheap networking for ultra-dense data centers.
\newblock In {\em HotNets-XIII Proceedings of the 13th ACM Workshop on Hot
  Topics in Networks}. ACM, October 2014.

\end{thebibliography}
\bibliographystyle{abbrv}

\end{document}